\documentclass[authoryear,review,3p,10pt,onecolumn,leqno]{elsarticle}
\usepackage[english]{babel}
\usepackage{epsfig, epstopdf}
\usepackage{color}
\usepackage{hyperref}
\usepackage{lineno}
\usepackage{amssymb,amsmath}
\usepackage{natbib}
\usepackage{setspace}
\usepackage{lscape}
\usepackage{afterpage,latexsym,epsfig,float}
\bibpunct{(}{)}{;}{a}{}{,}

\journal{Preprint version}

\begin{document}

\title{Credit allocation based on journal impact factor and\\ coauthorship contribution}

\author[IFOP,UTFS]{Javier E. Contreras-Reyes\corref{cor1}}
\ead{jecontrr@uc.cl, javier.contreras@ifop.cl}

\cortext[cor1]{Corresponding author, Phone +56 7 9608218. Address: Departamento de Econom\'ia y Estad\'istica, Divisi\'on de Investigaci\'on Pesquera, Instituto de Fomento Pesquero,
Blanco 839, Valpara\'iso, Chile.}
\address[IFOP]{Departamento de Econom\'ia y Estad\'istica, Divisi\'on de Investigaci\'on Pesquera,\\ Instituto de Fomento Pesquero (IFOP), Valpara\'iso, Chile}
\address[UTFS]{Departamento de Matem\'aticas, Universidad T\'ecnica Federico Santa Mar\'ia, Valpara\'iso, Chile}

\begin{abstract}
Some research institutions demand researchers to distribute the incomes they earn from publishing papers to their researchers and/or co-authors. In this study, we deal with the Impact Factor-based ranking journal as a criteria for the correct distribution of these incomes. We also include the Authorship Credit factor for distribution of the incomes among authors, using the geometric progression of Cantor's theory and the Harmonic Credit Index. Depending on the ranking of the journal, the proposed model develops a proper publication credit allocation among all authors. Moreover, our tool can be deployed in the evaluation of an institution for a funding program, as well as calculating the amounts necessary to incentivize research among personnel.\\
\end{abstract}

\maketitle

\section{Introduction}

Research institutions like universities or governmental/military institutes require staff to re-distribute the remunerations they receive from publishing in journals among their researchers/employers \citep{Groshen_1991}. Thus, one wage type earned for published papers serves as payment to co-authors. This problem is difficult to solve, given the prevalent conflicts of interest in many institutions, where distribution of payments is often unjust due to bad practice or ignorance. Therefore, a quantitative method to deal with distribution issues in relation to co-authorship is necessary.

The impact factor (IF) is a citation-based measure for performance related to prestige and proliferation of journals in which research institutions publish their papers \citep{Mattsson_et_al_2012}. A journal's IF gives the mean number of citations received by papers that have been published in these journals, and is one of the most popular indexes regarding quantitative methods to evaluate research \citep{Bouyssou_Marchant_2011}. Moreover, the number of citations determines the impact of journals on journal rankings \citep{Tsai_2014}. Generally, journals with a high IF contain a lower percentage of uncited articles \citep{Hsu_Huang_2012}.

In this paper, we present a model for the distribution of money benefits to an individual or a group of co-authors who published an International Scientific Indexing (ISI) paper. The model is based on two elements:

\begin{itemize}
  \item[(i)] the ranking of the journal based on Impact Factor (IF) as a criterion for weighing the distribution of these incomes; and 
  \item[(ii)] the authorship credit factor for a distribution of the incomes among the authors \citep{Lukovits_Vinkler_2012}, useful for multi-authored scientific publications.
\end{itemize}

Below our authorship factor considers the geometric progression of Cantor's theory \citep{Cantor_1883}, which we then compare with the harmonic credit approach of \citet{Hagen_2008}. Finally, we illustrate our results using three publications with varying numbers of co-authors, and journals with distinct IF and scholarly fields.\\

\section{Credit distribution model}

Let $t$ be the total publication credit allocated for one article ($t>0$), and let $r$ be the quotient of the journal's rank with respect to the total number of journals ($0<r<1$). The journal's rank is based on IF and is associated to the journal's area, as by the Web of Science. Let $p$ be a proportion arbitrarily assigned to an institution to define the income base from the total $t$. Thus, the total publication credit allocated for all authors is obtained by
\begin{equation}\label{CT}
Q_t(p,r) = pt + (1-p)(1-r)t,\quad 0\leq p\leq1,
\end{equation}

\begin{figure}[!htb]
\centering
\includegraphics[width=8cm,height=8cm]{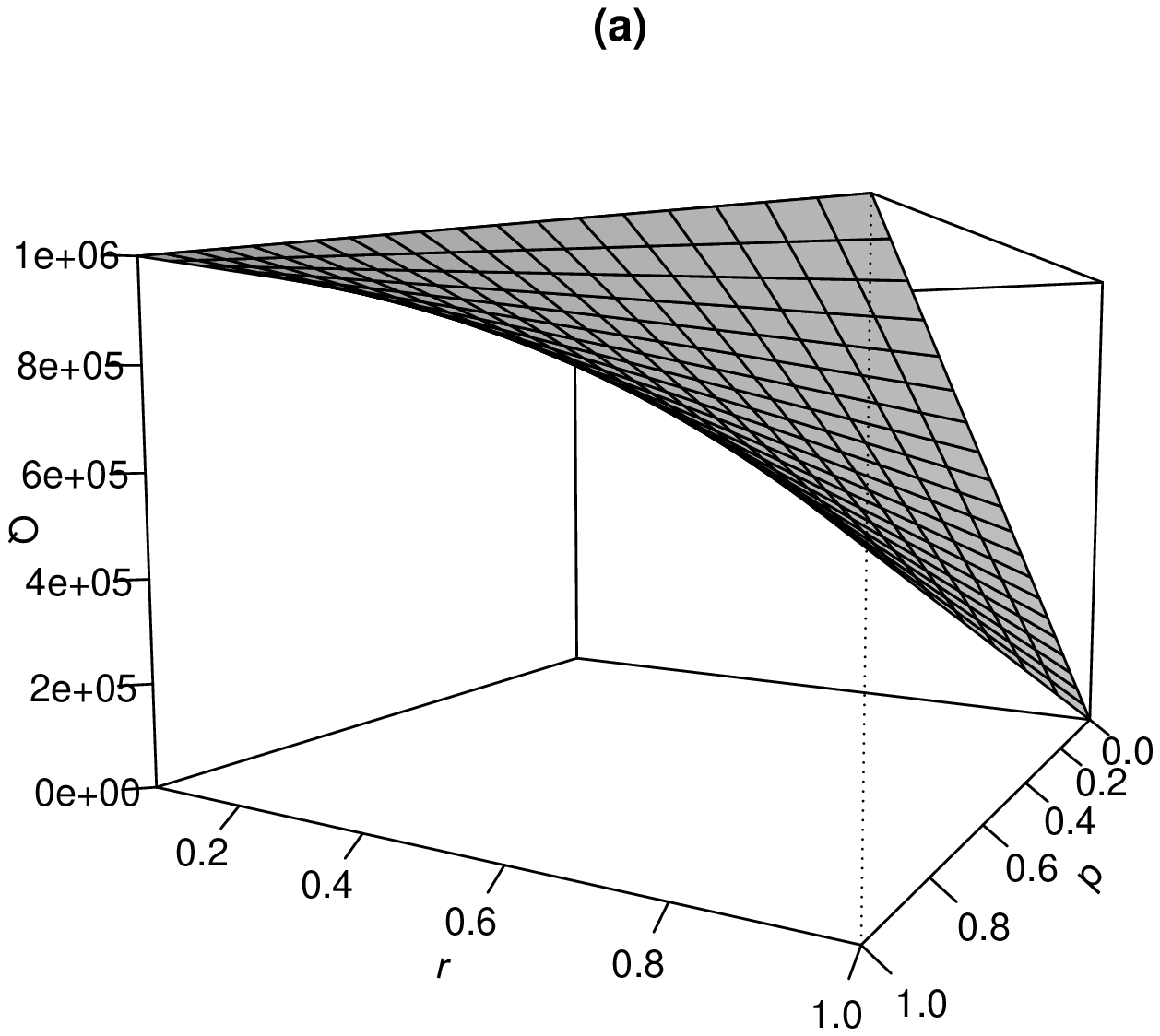}
\includegraphics[width=8cm,height=8cm]{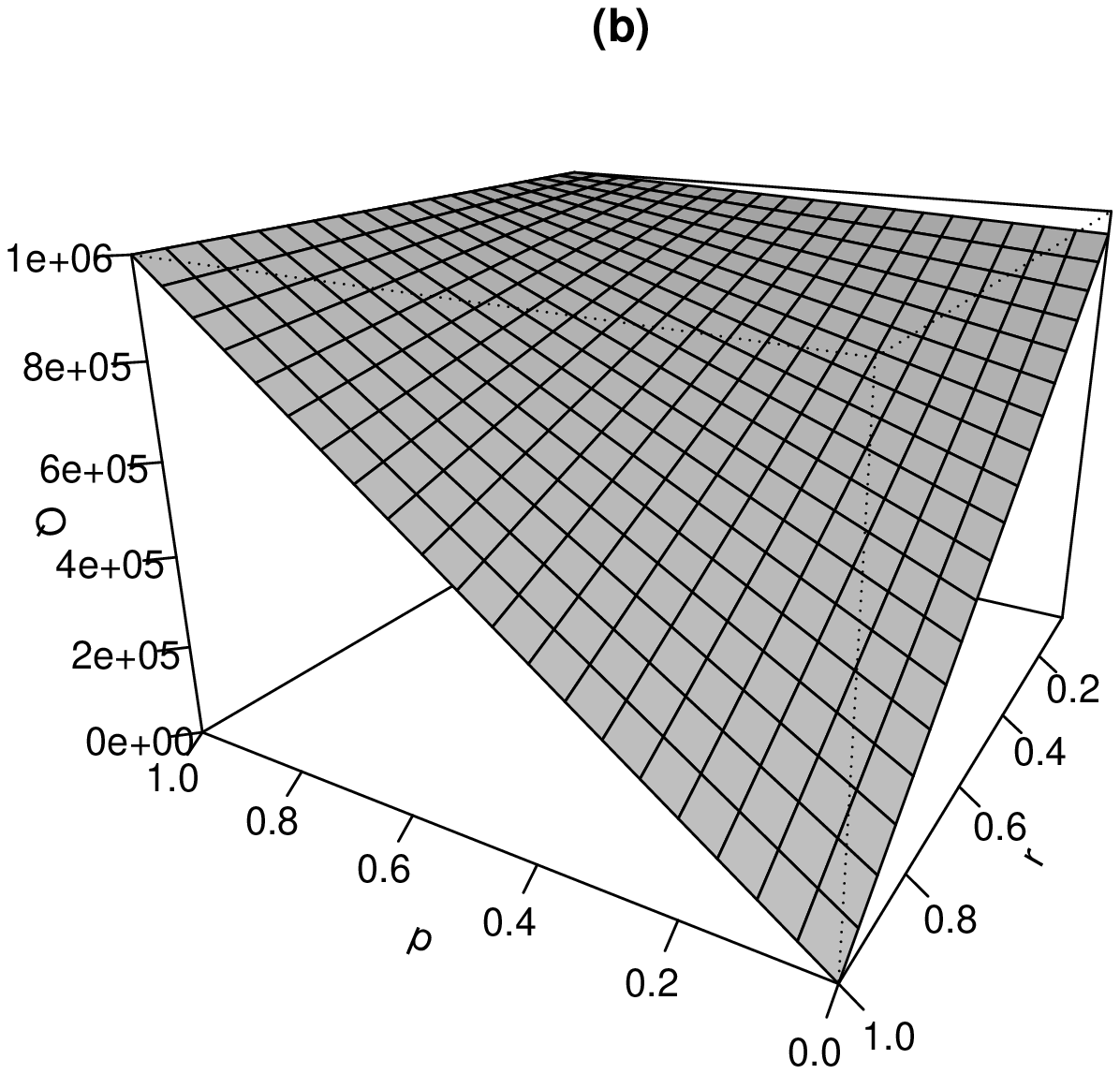}
\caption{Total publication credit $Q\equiv Q_t(p,r)$ allocated for all authors, with $t=10^{6}$, $0\leq p\leq1$, and $0< r\leq1$.}\label{G0}
\end{figure}

Function (\ref{CT}) depends only on the variables $p$ and $r$; $t$ is known. Evaluating point $(1,0)$ in the second derivative 
we obtain 
$$\frac{\partial^2}{\partial p\partial r}Q_t(p,r)=t>0,$$ 
i.e., $Q_t(p,r)$ reaches a relative maximum at $(1,0)$ and is given by the total available amount. Therefore, $Q_t(p,r)\leq t$ 
for all $0\leq p\leq1$ and for any $t>0$. We can interpret the model (\ref{CT}) with respect to $p$ as:
\begin{itemize}
  \item[(i)] if $p=1$, the credit $Q_t(1,r)$ corresponds to the available funding. Even if relative maximum $t$ is reached
  with this value, it does not incentivize the publication with respect to the journal quality;
  \item[(ii)] if $0<p<1$, the credit $Q_t(p,r)$ is the sum of a bonus base, $pt$, with an extra bonus, $(1-p)(1-r)t$,
  given by the IF of the journal. This is favourable if an institution wishes to incentivize a publication with respect to journal quality;
  \item[(iii)] if $p=0$, the credit $Q_t(0,r)$ depends not only on the available funding, but also relates to journal ranking. This is a 
  non-favourable case if an institution wishes to use all available funding to incentivize publication (Figure~\ref{G0} illustrates 
  the aforementioned cases).
\end{itemize}

When $r\approx0$ and $p=1$, the total amount is maximized, and is minimized when $r=1$ and $p\approx0$. For these values, the amount 
quickly decreases and the value $p=0.5$ could be acceptable to differentiate the credit base from the credit related to journal ranking. 
Values of $r$ near (and including it) 1 correspond to journals with the lowest impact, and $Q_t(0,r)=0$, $Q_t(p,r)=pt$ and $Q_t(1,r)=t$. 
This means that the ranking influences negatively the total publication credit. In a contrary case, values of $t$ near zero relate to 
the more prestigious journals. Given the values $p$, $r$ and $t$, model (\ref{CT}) gives the total credit to be allocated for the 
co-authors. As such, the institutions generally divided the credit equally among all co-authors \citep{Hagen_2008}.

In the next section, we consider three authorship credit indexes to distribute the credit $Q_t(p,r)$ in fractional form to privilege 
the main authors of a paper.\\

\section{Authorship credit indexes}

The total credit determined by model (\ref{CT}) is related to the publication; but how could it be assigned to the publication's authors? Authorship credit for multi-authored scientific publications is routinely allocated either by issuing full publication credit repeatedly to all co-authors \citep{Hagen_2008}, or by dividing one credit equally among all of them \citep{Karpov_2014}. We consider this latter option. That is, the credit allocated for a specific author is
\begin{equation}\label{CA}
A^i_t(p,r)=P_i\,Q_t(p,r),
\end{equation}
where $P_i$, $0<P_i<1$, corresponds to a succession of proportions or weight of the total credit. From (\ref{CA}) it becomes clear that infinite $P_i$ functions exist to share the incomes between $N$ authors. In addition, the condition called {\it sums of all shares $P_i$ is equal to 1} is accomplished \citep{Hagen_2008}. Among all of these functions proposed in the literature \citep{Egghe_et_al_2000,Hagen_2013}, we highlight the Harmonic Credit Index in the next subsection.\\

\subsection{Harmonic credit index}

\citet{Hagen_2008} proposed the Harmonic Credit Index (HCI) $H_i$ for the $i$th-author as follows:
\begin{eqnarray}
H_i&=&\frac{1}{i}\left(\sum_{j=1}^N \frac{1}{j}\right)^{-1},\quad i=1,\ldots,N,\label{HC}
\end{eqnarray}
where $N$ is the number of authors. The d'Alembert's ratio test for succession (\ref{HC}) shows that 
$$\lim_{i\rightarrow\infty}|H_{i+1}/H_i|=1,$$
meaning the test is inconclusive. However, the property (i) of Section~3.1, $H_1+\ldots+H_N=1$, ensures the convergence
of this succession \citep{Hagen_2008}. HCI ensures that: 
\begin{itemize}
  \item[(i)] the total publication credit is shared among all coauthors; 
  \item[(ii)] the main author gets most credit, and the $(i+1)$th author receives more credit than the $i$th author; 
  \item[(iii)] the higher the number of authors, the less credit per author. 
\end{itemize}  

In \citet{Hagen_2013}, HCI is compared with various co-author credit models such as fractional, Liu-Fang, Lukovits-Vinkler, and Trueba-Guerrero. 
For an empirical dataset including medicine, bibliometric literature, psychology, and chemistry \citep[see more details in ][]{Hagen_2010}, HCI 
performs better than its competitors explaining nearly 97\% of the variation versus, for example, 40\% of fractional credit index. 

The amount assigned to co-authors of (\ref{CA}) can be evaluated using (\ref{HC}), yielding
\begin{equation}\label{hci}
A^i_t(p,r)=H_i\,Q_t(p,r).
\end{equation}

Following from these properties of HCI, the total sum of $A^i_p(t,r)$ is equal to the credit $Q_t(p,r)$ assigned to all co-authors. Thus, 
formula (\ref{hci}) gives the complete HCI.\\

\subsection{Cantor's succession index}  

A geometric progression can be considered as a co-author credit index. The formula
\begin{eqnarray}
C_i&=&\frac{2^{i-1}}{3^i},\quad i=1,\ldots,N,\label{PG}
\end{eqnarray}
with $N$ the number of authors, corresponds to the proportion of the unit interval remaining, or Cantor's set \citep{Cantor_1883}.

The d'Alembert's ratio test shows that 
$$\lim_{i\rightarrow\infty}|C_{i+1}/C_i|=2/3<1,$$ 
i.e., the series converges absolutely.

\begin{figure}[!htb]
\centering
\includegraphics[width=10cm,height=10cm]{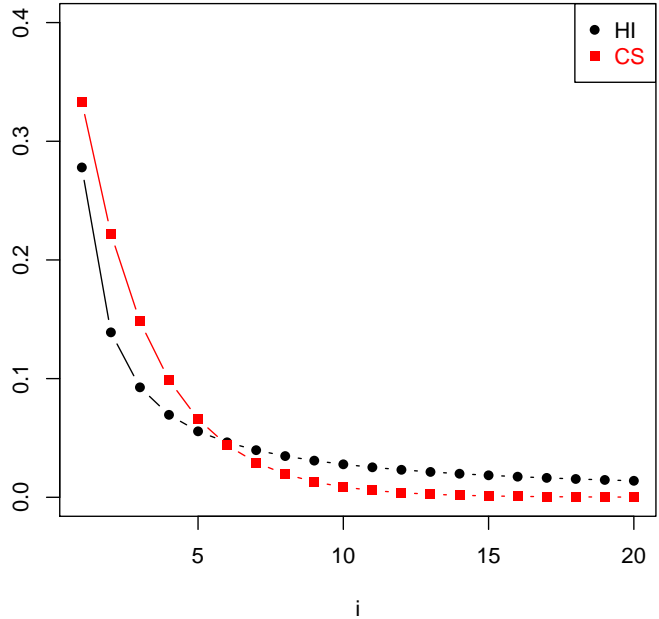}
\caption{Harmonic credit (HI) and Cantor's succession (CS) for $N=20$ authors.}\label{G1}
\end{figure}

Compared with $H_i$, $C_i$ do not depend on the total number of authors. However, the total sum of the $C_i$'s is less than 1; it is 1 only 
for a large number $N$ of authors. These series correspond to the total length removed from Cantor's sets. Figure~\ref{G1} compares both 
successions (\ref{HC}) and (\ref{PG}) between $N=20$ authors. It shows that the first five $H_i$ proportions differ from each author in a 
decreasing order. The first five $C_i$, however, are about similar to each author, but in decreasing order. Only for $i=6$ both successions are about equal 
and for $i>6$, $H_i$ tends to be larger than $C_i$ but with similar distribution among these authors. This illustrates that Cantor's succession 
is also a fractional counting, where one credit is divided non-uniformly among all co-authors \citep{Hagen_2008,Hagen_2010}. Therefore, this 
succession also corrects for the inflationary bias produced by multi-authored publications.

As in Section~3.1, co-author amount assignment of (\ref{CA}) can be evaluated using (\ref{PG}) to obtain
\begin{equation}\label{csi}
A^i_t(p,r)=C_i\,Q_t(p,r).
\end{equation}
Hereafter, we will refer to formula (\ref{csi}) as the Cantor's Succession Index (CSI). We see that the total sum of $A^i_p(t,r)$
is less than the total credit $Q_p(t,r)$. If a large number of co-authors worked on the publication, the total sum of $A^i_p(t,r)$
tends to be the total credit $Q_p(t,r)$. However, for a small number of authors, CSI produces an error in the distribution and a 
considerable bias, leading to ill-distributed credit. For this reason, and to obtain an index with which this important property 
is accomplished, it is necessary to implement a correction factor for CSI, as presented in the next subsection.\\

\subsection{Adjusted Cantor's succession index}

For a finite number of authors $N$, we have the publication credit
\begin{equation}\label{eps}
\varepsilon=Q_t(p,r)-\sum_{i=1}^N A^i_t(p,r) = Q_t(p,r)\left(1-\sum_{i=1}^N C_i\right)
\end{equation}
that is always positive and is zero when $N\rightarrow\infty$. Using (\ref{eps}), we define a corrected version
for $A^i_p(t,r)$ as
\begin{equation}\label{CAC}
{\bar A}^i_t(p,r) = A^i_t(p,r) + \frac{\varepsilon}{N}.
\end{equation}
Hereafter, we will refer to formula (\ref{CAC}) as the Adjusted Cantor's Succession Index (ACSI).

From (\ref{CAC}), it is clear that ${\bar A}^i_t(p,r)=A^i_t(p,r)$ when $N\rightarrow\infty$. Considering
(\ref{eps}) and (\ref{CAC}), we can corroborate that $\sum_{i=1}^{N}{\bar A}^i_t(p,r)=Q_t(p,r)$. Based
on ACSI, we obtain a new publication credit allocated for an specific $i$th author in (\ref{CA}) given by
\begin{equation}\label{CAC1}
{\bar A}^i_t(p,r)=C_i\,Q_t(p,r) + \frac{\varepsilon}{N}.
\end{equation}

ACSI also preserves the property of fractional counting, where one credit is divided non-uniformly and 
equally among the main co-authors and the rest, respectively.\\

\section{Examples}

We illustrate the performance of each co-author credit from ISI publications, considering formulas (\ref{hci}) and (\ref{CAC1}) and
a bonus base proportion of $p=0.5$, in the following examples:

\begin{enumerate}
\item The {\it Earth and Planetary Science Letters} ISI journal has an impact factor of 4.724 (according to the Web of Science 2013),
and ranked 5th out of 80 journals in the field of {\it Geochemistry \& Geophysics}. Considering research by \citet{Lange_et_al_2012},
for the values $t=2\times 10^6$, $r=0.063$ and $N=11$, the publication credit allocation between the $N$ authors is $Q_t(0.5, 0.063)=1,937,000$
for HCI and ACSI and $Q_t(0.5, 0.063)=1,914,606$ for CSI (see Table~\ref{T1}). The total proportion of $C_i$ is near 1 given the large
list of coauthors. This produces CSI near ACSI. However, for the properties mentioned in Section~3.1, HCI distributed the credit to the
three main authors in a non-uniform way, whereas ACSI provides more equal credits among them.

\begin{table}[h!]
\caption{Credit coauthorship distribution considering the paper \citet{Lange_et_al_2012}.}
\label{T1}
\begin{center}
\begin{tabular}{lrrrrr}
  \hline
    Author ($i$) & $C_i$ & CSI & ACSI & $H_i$ & HCI  \\
  \hline
1  & 0.333 & 645,666.67 & 647,702.46 & 0.331 & 641,416.78  \\
2  & 0.222 & 430,444.44 & 432,480.23 & 0.166 & 320,708.39  \\
3  & 0.148 & 286,962.96 & 288,998.75 & 0.110 & 213,805.59  \\
4  & 0.099 & 191,308.64 & 193,344.43 & 0.083 & 160,354.19  \\
5  & 0.066 & 127,539.10 & 129,574.89 & 0.066 & 128,283.36 \\
6  & 0.044 & 85,026.06 & 87,061.85 & 0.055 & 106,902.80 \\
7  & 0.029 & 56,684.04 & 58,719.83 & 0.047 & 91,630.97 \\
8  & 0.020 & 37,789.36 & 39,825.15 & 0.041 & 80,177.10 \\
9  & 0.013 & 25,192.91 & 27,228.70 & 0.037 & 71,268.53 \\
10  & 0.009 & 16,795.27 & 18,831.06 & 0.033 & 64,141.68 \\
11  & 0.006 & 11,196.85 & 13,232.64 & 0.030 & 58,310.62 \\
\hline
Total & 0.989 & 1,914,606.00 & 1,937,000.00 & 1 & 1,937,000.00 \\
  \hline
\end{tabular}
\end{center}
\end{table}

\item The {\it Fisheries Research} ISI journal has an impact factor of 1.843 (according to the Web of Science 2013), and ranked 12 out of
50 journals in the {\it Fisheries} field. Considering research by \citet{Contreras_et_al_2014}, for the values $t=1.2\times 10^6$,
$r=0.24$ and $N=3$, the publication credit allocation between the $N$ authors is $Q_t(0.5, 0.24)=1,056,000$ for HCI and ACSI and
$Q_t(0.5, 0.24)=743,111.1$ for CSI (see Table~\ref{T2}). In contrast to the first example, $C_i$ is far from 1, given the short list of
coauthors, thus CSI is far from ACSI. In this case, HCI should be more adequate to distribute the total amount between the three authors.
However, ACSI preserves the proportion among the authors.

\begin{table}[h!]
\caption{Credit coauthorship distribution considering the paper \citet{Contreras_et_al_2014}.}
\label{T2}
\begin{center}
\begin{tabular}{lrrrrr}
  \hline
    Author ($i$) & $C_i$ & CSI & ACSI & $H_i$ & HCI  \\
  \hline
1  & 0.333 & 352,000.0 & 456,296.3 & 0.545 & 576,000 \\
2  & 0.222 & 234,666.7 & 338,963.0 & 0.273 & 288,000 \\
3  & 0.148 & 156,444.4 & 260,740.7 & 0.182 & 192,000 \\
\hline
Total & 0.704 & 743,111.1 & 1,056,000.0 & 1 & 1,056,000 \\
  \hline
\end{tabular}
\end{center}
\end{table}

\item Consider now \citet{Ausloos_2015}. The {\it Physica A} ISI journal has an impact factor of 1.722 (according to the Web
of Science 2013), and is ranked 25 out of 78 journals in the {\it Physics Multidisciplinary} field. For the values $t=1.2\times 10^6$,
$r=0.321$ and $N=1$, the publication credit allocation for the author is $Q_t(0.5, 0.321)=792,308$ for HCI, CSI and ACSI. Given that we
have only one author, obviously CSI does not provide a precise criterion. Using the adjusted version, we obtain the same result for HCI.\\
\end{enumerate}

\section{Conclusions}

The proposed bonus distribution model gives a publication credit allocation associated to the performance of a journal, given by its IF-based ranking. This approach allows to restrict the credit allocated to each author, and so ranking criteria encourage research activity in an institution and to publish in higher-ranked journals. This tool is also helpful when an institution needs to be evaluated for a funding program as well as to determine where to direct amounts to incentivize research \citep{Egghe_et_al_2000}.

The proposed model is not restricted to a specific succession index. The most simple index is one that divides the total available amount equally among all co-authors \citep{Hagen_2008}. Cantor's index considers authorship rank instead, and the HCI considers these ranks and the number of co-authors. HCI provides different credits among the main authors, where this number depends on the total number of authors. However, the HCI should be employed for three main reasons:
\begin{itemize}
  \item[(i)] HCI proportions differ from each author in a decreasing order, allocating the higher amounts of credit to the prime authors, who make larger individual contributions to a paper \citep{Mattsson_et_al_2012};
  \item[(ii)] Adjusted Cantor's Index is still a fractional counting, and only divides non-uniformly the first fraction of the credit among all co-authors; and
  \item[(iii)] HCI's formula is much more tractable and simpler than ACSI. 
\end{itemize}

Since the selection of proportion $p$ of the proposed model is largely influenced by institution's policies, the model allows for sufficient flexibility to decide how to develop research among co-authors. Actually, several institutions such as universities choose option (i) of Section 2, assuming that an ISI paper deserves recognition only if it has the aforementioned indexation. This looks erroneous, however, because the journal ranking provides a qualification to measure its reputation with respect to an associated field  \citep{Bouyssou_Marchant_2011}.

Given that $H$-index is probably not the best indicator/predictor of the journal quality  \citep{Hirsch_2007}, the model can
be easily adapted to a more effective index to determine the ranking $r$. Additionally, further research using a stochastic 
$H$-index \citep{Nair_Turlach_2012} is needed, where the underlying proposed model is dependent on time.\\

\section*{Acknowledgements}

I would like to thank Paola Andrea Mar\'in (INE, Santiago, Chile) and Ghislaine Barr\'ia (IFOP, Valpara\'iso, Chile) for their helpful comments
which improve this work substantially.

\end{document}